\documentclass[letterpaper,USenglish]{lipics-v2016}
\usepackage{microtype}

\title{Toward Semantic Foundations for Program Editors
}
\titlerunning{Toward Semantic Foundations for Program Editors} 

\author[1]{Cyrus Omar}
\author[1]{Ian Voysey}
\author[2]{Michael Hilton}
\author[1]{Joshua Sunshine}
\author[1]{Claire Le Goues}
\author[1]{Jonathan Aldrich}
\author[3]{Matthew A. Hammer}
\affil[1]{Carnegie Mellon University, Pittsburgh, PA, USA\\
  \texttt{\{comar,iev,sunshine,clegoues,aldrich\}@cs.cmu.edu}}
\affil[2]{Oregon State University, Corvallis, OR, USA\\
\texttt{hiltonm@eecs.oregonstate.edu}}
\affil[3]{University of Colorado Boulder, Boulder, CO, USA\\
  \texttt{matthew.hammer@colorado.edu}}

\authorrunning{C. Omar, I. Voysey, M. Hilton, J. Sunshine, C. Le Goues, J. Aldrich, and M. Hammer} 

\Copyright{the authors}


\EventEditors{John Q. Open and Joan R. Access}
\EventNoEds{2}
\EventLongTitle{Summit oN Advances in Programming Languages (SNAPL 2017)}
\EventShortTitle{SNAPL 2017}
\EventAcronym{SNAPL}
\EventYear{2017}
\EventDate{May 7--10, 2017}
\EventLocation{Asilomar, California}
\EventLogo{}
\SeriesVolume{42}
\ArticleNo{23}

\usepackage{etoolbox}
\usepackage{hyperref}

\usepackage{listings}
\usepackage{graphicx}
\usepackage{wrapfig}
\usepackage{lmodern}
\usepackage{anyfontsize}
\usepackage{stmaryrd}
\SetSymbolFont{stmry}{bold}{U}{stmry}{m}{n}
\usepackage{amsmath,amsthm,amssymb}
\usepackage{thmtools,thm-restate}
\usepackage{todonotes}
\usepackage{enumerate}

\let\li\lstinline

\newcommand{\Hazel}[0]{\textsf{Hazel}}

\newcommand{\llparenthesiscolor}{\textcolor{violet}{\llparenthesis}}
\newcommand{\rrparenthesiscolor}{\textcolor{violet}{\rrparenthesis}}

\newcommand{\htau}{\dot{\tau}}
\newcommand{\tarr}[2]{#1 \rightarrow #2}
\newcommand{\tnum}{\texttt{num}}
\newcommand{\tehole}{\llparenthesiscolor\rrparenthesiscolor}

\newcommand{\hexp}{\dot{e}}
\newcommand{\hlam}[2]{\lambda #1.#2}
\newcommand{\hap}[2]{#1(#2)}
\newcommand{\hnum}[1]{\underline{#1}}
\newcommand{\hadd}[2]{#1 + #2}
\newcommand{\hehole}{\llparenthesiscolor\rrparenthesiscolor}
\newcommand{\hhole}[1]{\llparenthesiscolor#1\rrparenthesiscolor}

\begin{document}
\maketitle

\begin{abstract}
Programming language definitions assign formal meaning to {complete}
programs.
Programmers, however, spend a substantial amount of time interacting
with \emph{incomplete} programs -- programs with holes, type inconsistencies and binding inconsistencies -- using tools like program editors and
live programming environments (which interleave editing and
evaluation).
Semanticists have done comparatively little to formally characterize (1) the static and dynamic semantics of incomplete programs; (2) the 
actions available to programmers as they edit and inspect incomplete programs; and (3) the behavior of editor services that suggest likely edit actions to the programmer.

This paper serves as a vision statement for a research program that seeks to develop these ``missing'' semantic 
foundations. Our hope is that these contributions, which will take the form of a series of simple formal calculi equipped with a tractable metatheory, will guide the design of a variety of current and future interactive programming tools, much as various lambda calculi have guided modern language designs. Our own research will apply these principles in the design of \Hazel, an experimental \emph{live lab notebook} programming environment designed for data science tasks. We plan to co-design the \Hazel~language with the editor so that we can explore concepts such as edit-time semantic conflict resolution mechanisms and mechanisms that allow library providers to install library-specific editor services.

\end{abstract}

\section{Introduction}

Language-aware program editors (like Eclipse
or Emacs, with the appropriate  
extensions installed \cite{gamma2004contributing}) offer programmers a number of useful editor services. Simple examples include (1)
syntax highlighting, (2)
type inspection, (3)
navigation to variable binding sites, and (4)
refactoring services. More sophisticated editors provide context-aware code and action suggestions to the programmer (using various code completion, program synthesis and program repair techniques). Many editors also offer \emph{live programming}~\cite{McDirmid:2007:LUL:1297027.1297073,Burckhardt:2013:ACF:2491956.2462170} services, e.g. by displaying the run-time value of an expression directly within the editor as the program runs. 

When these editor services encounter \emph{complete programs} -- programs that are well-formed and semantically meaningful (i.e. assigned meaning) according to the definition of the language in use -- they can rely on a variety of well-understood reasoning principles and program manipulation techniques. For example, a syntax highlighter for well-formed programs can be generated automatically 
from a context-free grammar \cite{Brand:2001hl} and the remaining editor services enumerated above can follow the language's type and binding structure as specified by a standard static
semantics. Live programming services can additionally follow the language's dynamic semantics.

The problem, of course, is that many of the {edit states} encountered by a program editor do not correspond to complete programs. For example, the programmer may be in the midst of a transient edit, or the programmer may have introduced a type error somewhere in the program. Standard language definitions are silent about incomplete programs, so in these situations, simple program editors disable various editor services until the program is again in a complete state. In other words, useful editor services become unavailable when the programmer needs them most! More advanced editors attempt to continue to provide editor services during these incomplete states by using various \emph{ad hoc} and poorly understood heuristics that rely on idiosyncratic internal representations of  incomplete programs.

This paper advocates for a research program that seeks to understand both incomplete programs, and the editor services that interact with them, as semantically rich mathematical objects. This research program will broaden the scope of the ``programming language theory'' (PLT) tradition, which has made significant advances by treating complete programs, programming languages and logics as semantically rich mathematical objects.

In following the PLT tradition, we intend to start by developing a series of minimal calculi that build upon well-understood typed lambda calculi to capture the essential character of incomplete programs and various editor services of interest. Editor designers will be able to apply the insights gained from studying these calculi (together with insights gained from the study of human factors and other topics) to design more sophisticated program editors. Some of these editors will evolve directly from editors already in use today. In parallel with these efforts, we plan to design a ``clean-slate'' programming environment, \Hazel, based directly on these first principles. 
This will allow researchers to explore the frontier of what is possible when one considers languages and editors within a common theoretical framework. 
Such a clean-slate design will also likely prove useful in certain educational settings, and even some day evolve into a practical tool.

Figure \ref{fig:hazel-mockup} shows a mockup of the \Hazel~user interface, which is loosely modeled after the widely adopted IPython / Jupyter lab notebook interface~\cite{PER-GRA:2007}. 
This figure will serve as our running example 
throughout the remainder of the paper. Each section below briefly summarizes a fundamental problem that we must confront as we seek to develop a semantic foundation for advanced program editors. For each problem, we discuss existing approaches, including those advanced by our own recent research, and suggest a number of promising future research directions that we hope that the community will pursue.

\section{Problem 1: Syntactically Malformed Edit States} 
Textual program editors frequently encounter edit states
that are not well-formed with respect to the textual syntax of complete
programs. For example, consider a programmer constructing a call to a function \lstinline{std}: 
\[
\texttt{std(m, }
\]
There is a syntax error, so editor services that require a syntactically
complete program must be disabled. This is unsatisfying. 

Sophisticated editors like Eclipse, and editor generators like Spoofax \cite{DBLP:conf/oopsla/KatsV10}, use \emph{error recovery} heuristics that silently insert tokens so that the editor-internal representation is well-formed \cite{DBLP:journals/siamcomp/AhoP72,charles1991practical,graham1979practical,DBLP:conf/oopsla/KatsJNV09}. These heuristics are typically provided manually by the grammar designer, though certain heuristics can be generated semi-automatically by tools that are given a description of the scoping conventions of the language or of secondary notational conventions (e.g. whitespace) \cite{DBLP:conf/oopsla/KatsJNV09,DBLP:conf/sle/JongeNKV09}. Error recovery heuristics require guessing at the programmer's intent, so they are fundamentally \emph{ad hoc} and can confuse the programmer \cite{DBLP:conf/oopsla/KatsJNV09}.

\begin{figure}
\includegraphics[width=1.025\textwidth]{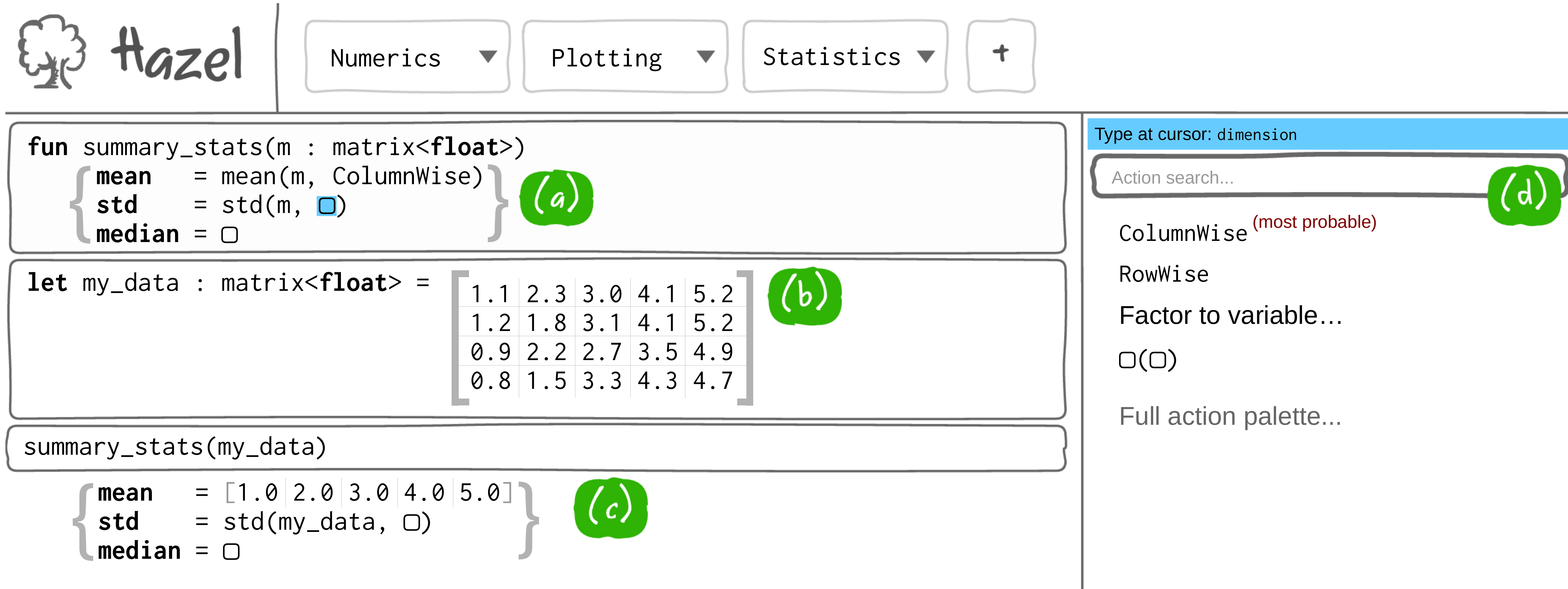}
\vspace{-5px}
\caption{A mockup of \Hazel.}
\vspace{-6px}
\label{fig:hazel-mockup}
\vspace{-6px}
\end{figure}

A more systematic alternative approach, and the approach that we plan to explore with \Hazel, is to build a
\emph{structure editor} -- a program editor where every edit state
maps onto a syntax tree, with \emph{holes} representing leaves of the tree
that have not yet been constructed.  This representation choice sidesteps the
problem of syntactically malformed edit states. Notice that in
Figure~\ref{fig:hazel-mockup}, the program fragment in cell
\textbf{(a)} contains holes, appearing as squares. This design also permits
non-textual \emph{projections} of expressions, e.g. 
the 2D projection of a matrix value in cell \textbf{(b)}.
We will return to the topic of non-textual projections below.

Structure editors have a long history. For example, the Cornell Program
Synthesizer was developed in the early 1980s \cite{teitelbaum_cornell_1981}. 
Although text-based syntax continues to predominate, there remains significant   
 interest in structure editors today, particularly in practice. For example,  Scratch is a 
structure editor that has achieved success as a tool for teaching children
how to program \cite{Resnick:2009:SP:1592761.1592779}. \texttt{mbeddr} is an editor for a C-like
language \cite{voelter_mbeddr:_2012}, built using the commercially supported MPS structure editor workbench \cite{voelter2011language}. TouchDevelop is an editor for an
object-oriented language \cite{tillmann_touchdevelop:_2011}. Lamdu \cite{lamdu} and Unison \cite{unison} are open source structure
editors for functional languages similar to Haskell. Most work on structure editors has focused on the user
interfaces that they present. This is important work -- presenting a
fluid user interface involving higher-level edit actions is a non-trivial
problem, and some aspects of this problem remain open even after many years of research. There is reason to be optimistic, however, with recent studies 
suggesting that programmers experienced with a modern keyboard-driven structure editor (e.g. \texttt{mbeddr}) 
can be highly productive \cite{DBLP:conf/vl/Asenov014,DBLP:conf/sle/VolterSBK14}.

Researchers have also explored various ``hybrid'' approaches, which incorporate holes into an otherwise textual program editor. These hybrid approaches are appealing in part because tools for interacting with text, like regular expressions and various differencing techniques used by version control systems, are already well-developed. For example, recent work on \emph{syntactic placeholders} envisions 
a textual program editor where edit actions cause textual placeholders (a.k.a. holes) of various sorts to appear, rather than leaving the program transiently malformed \cite{Amorim:2016:PSC:2997364.2997374}. This 
``approximates'' the experience of a structure editor in common usage, while allowing the programmer to perform arbitrary 
text edits when necessary. Some programming systems, e.g. recent iterations of the Glasgow Haskell Compiler (GHC) \cite{GCHWIKI} and the Agda proof assistant \cite{norell2009dependently}, support a workflow where the programmer places holes manually at locations in the program that remain under construction. Another hybrid approach would be to perform error recovery by attempting to insert holes into the internal representation used 
by the program editor, without including them in the surface syntax exposed to programmers.  If ``pure'' structure editing proves too rigid as we design \Hazel, we will explore hybrid approaches.

\section{Problem 2: Statically Meaningless Edit States} \label{sec:p-statics}

\begin{figure}[t]
$\arraycolsep=4pt\begin{array}{lllllll}
\mathsf{HTyp:} & \htau & ::= &
  \tarr{\htau}{\htau} ~\vert~
  \tnum ~\vert~
  \tehole
\\
\mathsf{HExp:} & \hexp & ::= &
  x ~\vert~
  \hlam{x}{\hexp} ~\vert~
  \hap{\hexp}{\hexp} ~\vert~
  \hnum{n} ~\vert~
  \hadd{\hexp}{\hexp} ~\vert~
  (\hexp : \htau) ~\vert~
  \hehole ~\vert~
  \hhole{\hexp}
\end{array}$
\caption{Syntax of H-types and H-expressions in the Hazelnut calculus \cite{popl-paper}.}
\label{fig:hexp-syntax}
\end{figure}

No matter how an 
editor confronts syntactically malformed edit states, it must also confront 
edit states that are syntactically well-formed but statically meaningless. For
example, the following value member definition (assuming an ML-like language) has a type inconsistency:
\begin{lstlisting}[numbers=none]
val x : float = std(m, ColumnWise)
\end{lstlisting}
because \li{std} has type \li{matrix(float) * dimension -> vec(float)},
but the type annotation on \li{x} is \li{float}, rather than \li{vec(float)}. This leaves the entire surrounding program
formally meaningless according to a standard static semantics.

In the presence of syntactic holes, the problem of reasoning statically about incomplete programs 
becomes even more interesting.  Consider the incomplete expression \texttt{std(m,~$\square$)} 
from cell \textbf{(a)} in Figure \ref{fig:hazel-mockup}.
Although it is intuitively apparent that the type of this expression, after hole instantiation, could only be \lstinline{vec(float)} (the return type of \lstinline{std}),
and that the hole must be instantiated with values of type \li{dimension}, the static
semantics of complete expressions is again silent about these matters. 

Various heuristic
approaches are implemented in Eclipse and other sophisticated tools, but the 
formal character of these heuristics are obscure, buried deep within their implementations. What is needed is a clear static semantics for incomplete programs, i.e. programs that contain holes (in both expressions and types), type inconsistencies, binding inconsistencies (i.e. unbound variables), and
other static problems. Such a static semantics is necessary for \Hazel~to be able to  provide type inspection services. For example, in the right column of Figure \ref{fig:hazel-mockup}, \Hazel~is informing the programmer that the expression at the cursor, highlighted in blue in cell \textbf{(a)}, must be of type \li{dimension}). Similarly, \Hazel~must be able to assign the incomplete function
\li{summary_stats} an incomplete function
type for it to be able to understand subsequent applications of \li{summary_stats}. Here, the function body has been filled out enough to be able to assign the function the following incomplete function type:
\[\texttt{matrix(float) -> \{ {mean} : vec(float), std : vec(float), median :~$\square$ \}}\] 

We have investigated a subset of this problem in recent work \cite{popl-paper} by defining a static
semantics for a simply typed lambda calculus (with a 
single base type, $\tnum$, for simplicity) extended with holes and type
inconsistencies (but no binding inconsistencies). Figure~\ref{fig:hexp-syntax} defines the syntactic
objects of this calculus -- \emph{H-types}, $\htau$,
are types with holes $\tehole$, and \emph{H-expressions}, $\hexp$, are
expressions with holes $\hehole$, and marked type inconsistencies,
$\hhole{\hexp}$. We call marked type inconsistencies \emph{non-empty holes},
because they mark portions of the syntax tree that remain
incomplete and behave semantically much like empty holes. Types and expressions that contain no holes are \emph{complete
  types} and \emph{complete expressions}, respectively.

 We will not reproduce further details here. Instead, let us simply note some interesting connections with other work. 

 First, type holes behave much like unknown types, $?$, from Siek and Taha's pioneering work on gradual typing \cite{Siek06a}. This discovery is quite encouraging, given that gradual typing is also motivated by a 
desire to make sense of one class of ``incomplete program'' -- programs that
have not been fully annotated with types.

Empty expression holes have also been studied formally, e.g. as the \emph{metavariables} 
of contextual modal type theory (CMTT) \cite{Nanevski2008}. In particular, expression
holes can have types and are surrounded by contexts, just as metavariables in
CMTT are associated with types and contexts. This begins to clarify the
logical meaning of a typing derivation in Hazelnut -- it conveys
well-typedness relative to an (implicit) modal context that extracts each
expression hole's type and context. The modal context must be emptied --
i.e. the expression holes must be instantiated with expressions of the
proper type in the proper context -- before the expression can be
considered complete. This relates to the notion of modal necessity in
contextual modal logic.

For interactive
proof assistants that support a tactic model based directly on hole
filling, the connection to CMTT and similar systems is quite salient. For
example, Beluga \cite{DBLP:conf/flops/Pientka10} is based on dependent CMTT
and aspects of Idris' editor support \cite{brady2013idris} are based on a similar system -- 
McBride's OLEG \cite{mcbride2000dependently}. 
As we will discuss in Sec. \ref{sec:actions}, our notion of a program editor supports actions beyond hole filling.

There are a number of future research directions that are worth exploring.

\vspace{-10px}
\subparagraph{Binding inconsistencies.} In the simple calculus developed so far,
    all variables must be bound before they are used,
    including those in holes. We plan extend Hazelnut to support reasoning
    when a variable is mentioned without having been bound (as is a common workflow). Dagenais and
    Hendren also studied how to reason statically about programs
    with binding errors using a constraint system, focusing on
    Java programs whose imports are not completely known~\cite{DBLP:conf/oopsla/DagenaisH08}. They neither
    considered programs with holes or other type inconsistencies,
    nor did they formally specify their
    technique. However, they provide a useful starting point.

\vspace{-10px}
\subparagraph{Expressiveness.} The simple calculus discussed above 
    is only as expressive as the typed lambda calculus with numbers. We must scale up the semantics to handle other modern language
    features. Our plan is to focus initially on functional language
    constructs (so that \Hazel ~can be used to teach courses that
    are today taught using Standard ML, OCaml or Haskell). This will include recursive and
    polymorphic functions, recursive types, and labeled product (record) and sum types.
    We also propose to investigate ML-style structural pattern
    matching. All of these will require defining new sorts of holes and static
    inconsistencies, including: (1) non-empty holes at the type level, to handle
    kind inconsistencies; (2) holes in label position; and (3) holes and type inconsistencies in patterns. 

\vspace{-10px}
\subparagraph{Automation.} Although we plan to
    explore some of these language extensions 
    ``manually,'' extending our existing mechanized metatheory, we ultimately plan 
    to \emph{automatically}
    generate a statics for incomplete terms from a standard statics for complete terms,
    annotated perhaps with additional information. There is some precedent for
    this in recent work on the Gradualizer, which is capable of
    producing a gradual type system from a standard type system with lightweight
    annotations that communicate the intended polarities of certain
    constructs~\cite{DBLP:conf/popl/CiminiS16}. However, although it provides a good starting point, gradual type systems 
    only consider the problem of holes in 
    types.
    Our plan is to build 
    upon existing proof automation techniques, e.g. Agda's reflection \cite{van2012engineering} (in part because our present mechanization effort is in Agda).

\section{Problem 3: Dynamically Meaningless Edit States} Modern programming
tools are increasingly moving beyond simple ``batch'' programming models by
incorporating \emph{live programming} features that interleave editing and
evaluation \cite{DBLP:conf/icse/Tanimoto13,DBLP:journals/vlc/Tanimoto90,McDirmid:2007:LUL:1297027.1297073}. These tools provide programmers with rapid feedback about the
dynamic behavior of the program they are editing, or selected portions thereof \cite{McDirmid:2013:ULP:2509578.2509585}. Examples include \emph{lab notebooks},
e.g. the popular IPython/Jupyter~\cite{PER-GRA:2007}, which allow the
programmer to interactively edit and evaluate program fragments organized into a
sequence of cells (an extension of the read-eval-print loop (REPL)); spreadsheets; {live graphics programming environments}, e.g. SuperGlue \cite{McDirmid:2007:LUL:1297027.1297073}, Sketch-n-Sketch \cite{DBLP:conf/pldi/ChughHSA16} and the tools demonstrated by Bret Victor in his lectures \cite{victor2012inventing}; the TouchDevelop live UI framework \cite{Burckhardt:2013:ACF:2491956.2462170}; and live visual and auditory dataflow languages \cite{DBLP:conf/vl/BurnettAW98}. In the words of Burckhardt et al. \cite{Burckhardt:2013:ACF:2491956.2462170}, live programming environments 
``capture the imagination of today's programmers and promise to narrow the temporal and perceptive gap 
between program development and code execution''. 

Our proposed design for \Hazel~combines aspects of several of these designs to form a \emph{live lab notebook interface}. 
It will use the edit state of each cell to continuously update the output
value displayed for that cell and subsequent cells that depend on
it. Uniquely, rather than providing meaningful feedback about the dynamic
behavior only once a cell becomes complete, \Hazel~will provide meaningful feedback also
about the dynamic behavior of incomplete cells (and thereby further tighten Burckhardt's ``perceptive gap'').

For example, in cell \textbf{(c)} of Figure~\ref{fig:hazel-mockup}, the
programmer applies  the incomplete function \li{summary_stats} to 
the matrix \lstinline{my_data}, and 
the editor is still able to display a result.
The value of the column-wise mean is fully determined, because evaluation does
not encounter any holes, whereas the standard deviation and median computations
cannot be fully evaluated. Notice, however, that the standard
deviation computation does communicate the substitution of the applied argument,
\li{my_data}, for the variable \li{m}.\footnote{To avoid exposing the internals
of imported library functions, evaluation does not step into functions, like
\li{std}, that have been imported from external libraries indicated by the row at the top of Figure \ref{fig:hazel-mockup} (unless specifically
requested, not shown).}

To realize this functionality, we need a
{dynamic semantics for incomplete programs} that builds upon our proposed
static semantics. There is some precedent for this: research in gradual typing
considers the dynamic semantics of programs with holes in types, and our
proposed static semantics for incomplete programs borrows technical machinery
from theoretical work on gradual typing~\cite{Siek06a}. However, we need a dynamic semantics for 
 incomplete programs that also have expression holes (and in the future, other sorts of holes). 

Research on CMTT has not yet 
 considered the problem of evaluating expressions under a non-empty metavariable context. 
Normally, this would violate the classical notion of Progress -- 
evaluation can neither proceed, nor has it produced a value. We conjecture that this is
resolved by (1) positively characterizing \emph{indeterminate} 
evaluation states, those where a hole blocks progress at all locations
within the expression, and (2) defining
a notion of Indeterminate Progress that allows for evaluation to stop at an 
indeterminate evaluation state. By gradualizing CMTT and defining these notions, we believe we can achieve the basic functionality described above.

There are several more applications that we aim to explore after developing these initial foundations. For example, it would be useful for the programmer to be able to select a hole that appears in an indeterminate state and be taken to its original location. There, they should be able to inspect the \emph{value} of a subexpression under the cursor in the environment of the selected hole (rather than just its type). Again, CMTT's closures provide a theoretical starting point for this debugger service. 

It would also be useful to be able to continue evaluation where it left off after making an edit to the program that corresponds to hole instantiation. This would require proving a commutativity property regarding hole instantiation. Fortunately, initial research on commutativity properties for holes has been conducted for CMTT, which will serve as a starting point for this work \cite{Nanevski2008}. There are likely to be interesting new theoretical questions (and, likely, some limitations) that arise if one adds non-termination and memory effects. 

Relatedly, IPython/Jupyter \cite{PER-GRA:2007} support a feature whereby numeric variable(s) in cells can be marked as being ``interactive'', which causes the user interface to display a slider. As the slider value changes, the value of the cell is recomputed. It would be useful to be able to use the mechanisms just proposed to incrementalize parts of this recomputation.

       \section{Problem 4: A Calculus of Edit Actions}\label{sec:actions} The previous sections 
considered the structure and meaning of intermediate edit states. However, to
understand the act of \emph{editing} itself, we need \emph{a calculus of edit actions} that governs transitions between these edit
states. 

In a structure editor, the ideal would be for every possible edit state to be both statically and
dynamically meaningful according to the semantics proposed in the previous two
sections. This corresponds formally to proving a
metatheorem about the action semantics: when the initial edit state is
semantically meaningful, the edit state that results from performing an action
is as well. In a textual or hybrid setting, these structured edit actions would need to be supplemented by lower-level text edit actions that may not maintain this invariant. In addition to this crucial metatheorem, which
we call \emph{sensibility}, there are a number of other metatheorems of interest
that establish the expressive power of the action semantics, e.g. that every well-typed term can be constructed by some sequence of edit actions. 

In our recent work on Hazelnut, we have developed an action calculus
for the minimal calculus of H-types and H-expressions described in
Section~\ref{sec:p-statics} \cite{popl-paper}. We have mechanically proven the sensibility invariant, as well as expressivity metatheorems, using the Agda proof assistant. What remains is to investigate \emph{action composition principles}. For example, it would be worthwhile to investigate the 
notion of an \emph{action macro}, whereby 
functional programs could themselves be lifted to the level of actions to
compute non-trivial compound actions. Such compound actions would give a
uniform description of transformations ranging from the simple---like
``move the cursor to the next hole to the right''---to quite complex whole
program refactorings, while r{}emaining subject to the core semantics. Using proof automation, it should be possible to prove that an action macro implements derived action logic that 
is admissible with respect to the core semantics. This would eliminate the possibility of ``edit-time'' errors. This is closely related to work on tactic languages in proof assistants, e.g. the Mtac typed macro language for Coq \cite{ziliani2015mtac}, differing again in that the action language involves notions other than hole filling. 

\section{Problem 5: Meaningful Suggestion Generation and Ranking} 
The simplest 
edit actions will be bound to keyboard shortcuts. However, \Hazel~will also provide suggestions to help the programmer edit incomplete
programs by providing a \emph{suggestion palette}, marked \textbf{(d)} in
Figure~\ref{fig:hazel-mockup}.  This palette will suggest semantically
relevant code  
snippets when the cursor is on an empty hole. It will also suggest other relevant
edit actions, including high-level edit actions implemented by imported action
macros (e.g. the refactoring action in Figure~\ref{fig:hazel-mockup}).  When the
cursor is on a non-empty hole, indicating a static error, it will suggest
bug fixes. We plan to also consider bugs that do not correspond to static
errors, including those identified explicitly by the programmer, and those
related to assertion failures or exceptions encountered when using the live
programming features of \Hazel. In these situations, we plan to build on 
existing automated fault localization techniques~\cite{Jones02,
  Qi13issta,Renieris03}.

Note that features like these are not themselves novel. Many editors provide
contextually relevant suggestions. Indeed, suggestion generation is
closely related to several major research areas: code
completion~\cite{Muslu12icse-nier,icse-naturalness12}, program
synthesis~\cite{Gulwani2010}, and program
repair~\cite{legoues12tse,angelix,prophet,Ke15ase}. 

The problems that such existing systems encounter is exactly the problem we have 
been discussing throughout this proposal: when attempting to integrate these 
features into an editor, it is difficult to reason about malformed or meaningless
edit states. Many of these systems therefore fall back onto tokenized representations of programs \cite{icse-naturalness12}. Because \Hazel~ will maintain the invariant that every
edit state is a syntactically and semantically meaningful formal structure, we can develop a
more principled solution to the problem of generating meaningful suggestions. In particular,
we will be able to \emph{prove} that every action suggestion generated for a particular edit state is 
meaningful for that edit state.

In addition to investigating the problem of populating the suggestion palette
with semantically valid actions, we will consider the problem of evaluating
the statistical likelihood of the suggestions. This
requires developing a statistical model over actions.  We will prove that this statistical model is a
proper probability distribution (e.g. that it ``integrates'' to 1), and that it
assigns zero probability to semantically invalid actions. We will also develop techniques for estimating the parameters of these distributions from a corpus of code or a corpus of edit actions. Collectively, we refer 
to these contributions as a \emph{statistical action suggestion semantics}. 

Ultimately, we envision this work as being the foundation for an \emph{intelligent programmer's assistant} that is able to integrate semantic information gathered from the incomplete program with statistics gathered from other programs and interactions that the system has observed to do much of the ``tedious'' labor of programming, without hiding the generated code from the programmer (as is the case with fully automated program synthesis techniques). 

\section{Language-Editor Co-Design}

In designing \Hazel, we are intentionally blurring the line between the programming language and the program editor. This opens up a number of interesting research directions in {language-editor co-design}. For example, it may be possible to recast  ``tricky'' language mechanisms, like function overloading, type classes \cite{hall96:_typeclasses}, implicit values, and unqualified imports, as editor mechanisms. Because we will be treating programming as a structured conversation between the programmer and the programming environment, the editor can simply ask the programmer to resolve ambiguities when they arise. The programmer's choice is then stored unambiguously in the underlying syntax tree.

Another important research
direction lies in exploring how types can be used to control  
the presentation of expressions in the editor. In the textual setting, we have developed \emph{type-specific
languages} (TSLs) \cite{TSLs}. It should be possible to define an analagous notion of \emph{type-specific projections} (TSPs) in the setting of a structure editor. For example, the matrix projection shown in Figure \ref{fig:hazel-mockup} need not be built in to \Hazel. Instead, the \li{Numerics} library provider will be able to introduce this logic. In particular, TSPs will define not only derived visual forms, but also derived edit actions (e.g. ``add new column'' for the example just given.) It should be possible to switch between multiple projections (including purely textual projections) while editing code and interacting with values. This line of research is also related to our work on \emph{active code completion}, which investigated type-specific code generation interfaces in a textual program editor (Eclipse) \cite{Omar:2012:ACC:2337223.2337324}. 

Another interesting direction is that of {semantic, interactive documentation}. In particular, in \Hazel, references to program structures that appear in documentation will be treated in the same way as other references and be subject to renaming and other operations. Documentation will also be capable of containing expressions of arbitrary types (e.g. of the \li{Image} or \li{Link} type). Together with the type-specific projection mechanism mentioned above, we hope that this will allow \Hazel~ to function not only as a structured programming environment, but also as a {structured document authoring environment}! By understanding hyperlinks as variable references (in, perhaps, a different modality \cite{vii2007type}), we may be able to blur the line between a module and a webpage. 

\section{Conclusion}
To summarize, there are a number of interesting semantic questions that come up in the design of program editors. We advocate a research program that studies these problems using mathematical tools previously used to study programming languages and complete programs. This work will both demystify the design of program editors and open up the doors for a number of advanced editor services. Ultimately, we envision an intelligent programmer's assistant that combines a deep semantic understanding of incomplete programs with a broad statistical understanding of common idioms to help humans author both programs and documents (as one and the same sort of artifact.)

\section*{Acknowledgments}

We thank the SNAPL~2017 reviewers and our paper shepherd Nate Foster for the thoughtful comments and suggestions. 
This work is supported in part through a gift from
Mozilla; by NSF under grant 
numbers CCF-1619282, 1553741 and 1439957; by AFRL and DARPA under agreement \#FA8750-16-2-0042; and by the NSA under lablet contract \#H98230-14-C-0140.  
Any opinions, findings, and conclusions or recommendations expressed
in this material are those of the author(s) and do not necessarily
reflect the views of Mozilla, NSF, AFRL, DARPA or NSA.

\bibliographystyle{plainurl}

\bibliography{Omar}

\end{document}